\documentclass[final,times,twocolumn]{elsarticle}
\usepackage{graphicx,amssymb,amsmath}
\usepackage{multirow,dcolumn,bm,latexsym,soul,nicefrac}
\newcolumntype{K}[1]{>{\centering\arraybackslash}p{#1}}

\journal{Physics Letters B}
\begin{document}
\newcommand{\be}{\begin{equation}}
\newcommand{\ee}{\end{equation}}
\newcommand{\bq}{\begin{eqnarray}}
\newcommand{\eq}{\end{eqnarray}}

\begin{frontmatter}

\title{Fisher matrix forecasts for astrophysical tests of the stability of the fine-structure constant}
\author[inst1,inst2]{C. S. Alves}\ead{up201403516@fc.up.pt}
\author[inst1,inst2]{T. A. Silva}\ead{up201405824@fc.up.pt}
\author[inst1,inst3]{C. J. A. P. Martins\corref{cor1}}\ead{Carlos.Martins@astro.up.pt}
\author[inst1,inst2,inst3]{A. C. O. Leite}\ead{Ana.Leite@astro.up.pt}
\address[inst1]{Centro de Astrof\'{\i}sica, Universidade do Porto, Rua das Estrelas, 4150-762 Porto, Portugal}
\address[inst2]{Faculdade de Ci\^encias, Universidade do Porto, Rua do Campo Alegre 687, 4169-007 Porto, Portugal}
\address[inst3]{Instituto de Astrof\'{\i}sica e Ci\^encias do Espa\c co, CAUP, Rua das Estrelas, 4150-762 Porto, Portugal}
\cortext[cor1]{Corresponding author}

\begin{abstract}
We use Fisher Matrix analysis techniques to forecast the cosmological impact of astrophysical tests of the stability of the fine-structure constant to be carried out by the forthcoming ESPRESSO spectrograph at the VLT (due for commissioning in late 2017), as well by the planned high-resolution spectrograph (currently in Phase A) for the European Extremely Large Telescope. Assuming a fiducial model without $\alpha$ variations, we show that ESPRESSO can improve current bounds on the E\"{o}tv\"{o}s parameter---which quantifies Weak Equivalence Principle violations---by up to two orders of magnitude, leading to stronger bounds than those expected from the ongoing tests with the MICROSCOPE satellite, while constraints from the E-ELT should be competitive with those of the proposed STEP satellite. Should an $\alpha$ variation be detected, these measurements will further constrain cosmological parameters, being particularly sensitive to the dynamics of dark energy.
\end{abstract}


\end{frontmatter}

\section{Introduction}

Astrophysical tests of the stability of fundamental couplings are an extremely active area of observational research \cite{Uzan,grg}. The deep conceptual importance of carrying out these tests has been complemented by recent (even if somewhat controversial \cite{Syst}) evidence for such a variation \cite{Dipole}, coming from high-resolution optical/UV spectroscopic measurements of the fine-structure constant $\alpha$ in absorption systems along the line of sight of bright quasars. The forthcoming ESPRESSO spectrograph \cite{ESPRESSO}, due for commissioning at the combined Coud\'e focus of ESO's VLT in late 2017, should significantly improve the sensitivity of these tests, as well as the degree of control over possible systematics.

Moreover, the results of these tests---whether they are detections of variations or null results---have a range of additional cosmological implications. They provide competitive constraints on Weak Equivalence Principle (WEP) violations \cite{Uzan,Pinho2,Bek} and, in the more natural scenarios where the same dynamical degree of freedom is responsible both for the dark energy and the $\alpha$ variation, can also be used in combination with standard cosmological observables to constrain the dark energy equation of state \cite{Pinho1,Pinho3} and indeed to reconstruct its redshift dependence \cite{Amendola,Leite2}.

While current data already provides useful constraints, the imminent availability of more precise measurements from the ESPRESSO spectrograph will have a significant impact in the field. In this work we apply standard Fisher Matrix analysis techniques to forecast the improvements that may be expected from ESPRESSO, but we also take the opportunity to look further ahead and discuss additional gains in sensitivity from the European Extremely Large Telescope (E-ELT), whose first light will be in 2024.

\section{Varying $\alpha$, dark energy and the Weak Equivalence principle}

Dynamical scalar fields in an effective four-dimensional field theory are naturally expected to couple to the rest of the theory, unless a (still unknown) symmetry is postulated to suppress these couplings \cite{Carroll,Dvali,Chiba}. We will assume that this coupling does exist for the dynamical degree of freedom responsible for the dark energy, assumed to be a dynamical scalar field denoted $\phi$. Specifically the coupling to the electromagnetic sector is due to a gauge kinetic function $B_F(\phi)$
\begin{equation}
{\cal L}_{\phi F} = - \frac{1}{4} B_F(\phi) F_{\mu\nu}F^{\mu\nu}\,.
\end{equation}
This function can be assumed to be linear,
\begin{equation}
B_F(\phi) = 1- \zeta \kappa (\phi-\phi_0)\,,
\end{equation}
(where $\kappa^2=8\pi G$) since, as has been pointed out in \cite{Dvali}, the absence of such a term would require the presence of a $\phi\to-\phi$ symmetry, but such a symmetry must be broken throughout most of the cosmological evolution. The dimensionless parameter $\zeta$ quantifies the strength of the coupling. With these assumptions one can explicitly relate the evolution of $\alpha$ to that of dark energy \cite{Pinho2,Erminia2}. The evolution of $\alpha$ can be written
\begin{equation}
\frac{\Delta \alpha}{\alpha} \equiv \frac{\alpha-\alpha_0}{\alpha_0} =B_F^{-1}(\phi)-1=
\zeta \kappa (\phi-\phi_0) \,,
\end{equation}
and defining the fraction of the dark energy density (the ratio of the energy density of the scalar field to the total energy density, which also includes a matter component) as a function of redshift $z$ as follows
\begin{equation}
\Omega_\phi (z) \equiv \frac{\rho_\phi(z)}{\rho_{\rm tot}(z)} \simeq \frac{\rho_\phi(z)}{\rho_\phi(z)+\rho_m(z)} \,,
\end{equation}
where in the last step we have neglected the contribution from the radiation density (we will be interested in low redshifts, $z<5$, where it is indeed negligible), the evolution of the scalar field can be expressed in terms of $\Omega_\phi$ and of the dark energy equation of state $w_\phi$ as
\begin{equation}\label{dynphi}
1+w_\phi = \frac{(\kappa\phi')^2}{3 \Omega_\phi} \,,
\end{equation}
with the prime denoting the derivative with respect to the logarithm of the scale factor. Putting the two together we finally obtain
\begin{equation} \label{eq:dalfa}
\frac{\Delta\alpha}{\alpha}(z) =\zeta \int_0^{z}\sqrt{3\Omega_\phi(z')\left[1+w_\phi(z')\right]}\frac{dz'}{1+z'}\,.
\end{equation}
The above relation assumes a canonical scalar field, but the argument can be repeated for phantom fields, leading to
\begin{equation} \label{eq:dalfa2}
\frac{\Delta\alpha}{\alpha}(z) =-\zeta \int_0^{z}\sqrt{3\Omega_\phi(z')\left|1+w_\phi(z')\right|}\frac{dz'}{1+z'}\,;
\end{equation}
the change of sign stems from the fact that one expects phantom fields to roll up the potential rather than down. Note that in these models the evolution of $\alpha$ can be expressed as a function of cosmological parameters plus the coupling $\zeta$, without explicit reference to the putative underlying scalar field. In these models the proton and neutron masses are also expected to vary---by different amounts---due to the electromagnetic corrections of their masses. Therefore, local tests of the Equivalence Principle also constrain the dimensionless coupling parameter $\zeta$ \cite{Uzan}, and (more to the point for our present purposes) they provide us with a prior on it.

We note that there is in principle an additional source term driving the evolution of the scalar field, due to the derivative of the gauge kinetic function, i.e. a term proportional to $F^2B_F'$. By comparison to the standard (kinetic and potential energy) terms, the contribution of this term is expected to be subdominant, both because its average is zero for a radiation fluid and because the corresponding term for the baryonic density is constrained by the aforementioned Equivalence Principle tests. For these reasons, in what follows we neglect this term (which would lead to environmental dependencies). We nevertheless note that this term can play a role in scenarios where the dominant standard term is suppressed.

A light scalar field such as we are considering inevitably couples to nucleons due to the $\alpha$ dependence of their masses, and therefore it mediates an isotope-dependent long-range force. This can be quantified through the dimensionless E\"{o}tv\"{o}s parameter $\eta$, which describes the level of violation of the WEP \cite{Uzan}. One can show that for the class of models we are considering the E\"{o}tv\"{o}s parameter and the dimensionless coupling $\zeta$ are simply related by \cite{Uzan,Dvali,Chiba}
\begin{equation} \label{eq:eotvos}
\eta \approx 10^{-3}\zeta^2\,;
\end{equation}
we note that while this relation is correct for the simplest canonical scalar field models we will consider in what follows, it is somewhat model-dependent (for example, it is linear rather than quadratic in $\zeta$ for Bekenstein-type models \cite{Bek}).

\section{Forecasting tools and fiducial models}

We will be considering three fiducial dynamical dark energy models where the scalar field also leads to $\alpha$ variations according to Eq. \ref{eq:dalfa}, as follows
\begin{itemize}
\item A constant dark energy equation of state, $w_0=const.$
\item A dilaton-type model where the scalar field $\phi$ behaves as $\phi(z)\propto(1+z)$; this is well motivated in string theory inspired models \cite{DPV}, but for our purposes it also has the advantage that despite the fact that it leads to a relatively complicated dark energy equation of state
\begin{equation}\label{darkeos2}
w(z)=\frac{[1-\Omega_\phi(1+w_0)]w_0}{\Omega_m(1+w_0)(1+z)^{3[1-\Omega_\phi(1+w_0)]}-w_0}\,,
\end{equation}
(where we are assuming flat universes, so the present-day values of the matter and dark energy fractions satisfy $\Omega_m+\Omega_\phi=1$); in this case Eq. \ref{eq:dalfa} simplifies to \cite{Pinho2}
\begin{equation}\label{dilalpha}
\frac{\Delta\alpha}{\alpha}(z)=\zeta\, \sqrt{3\Omega_\phi(1+w_0)}\, \ln{(1+z)}\,.
\end{equation}
Thus this case allows us to carry out analytic calculations, which we have used to validate our numerical pipeline.
\item The well-known Chevallier-Polarski-Linder (CPL) parametrization \cite{CPL1,CPL2}, where the redshift dependence of the dark energy equation of state is described by two separate parameters, $w_0$ (which is still its present-day value) and $w_a$ describing its evolution, as follows
\be
w(z)=w_0+w_a\frac{z}{1+z}\,.
\ee
\end{itemize}
All of these have been used in previous works to obtain constraints from current data \cite{Pinho2,Pinho1,Pinho3} or to forecast dark energy equation of state reconstructions \cite{Amendola,Leite2}, and therefore these previous works can easily be compared with ours.

Our forecasts were done with a Fisher Matrix analysis \cite{FMA1,FMA2}. If we have a set of M model parameters $(p_1, p_2, ..., p_M)$ and N observables---that is, measured quantities---$(f_1, f_2, ..., f_N)$ with uncertainties $(\sigma_1, \sigma_2, ..., \sigma_N)$, then the Fisher matrix is
\be
F_{ij}=\sum_{a=1}^N\frac{\partial f_a}{\partial p_i}\frac{1}{\sigma^2_a}\frac{\partial f_a}{\partial p_j}\,.
\ee
For an unbiased estimator, if we don't marginalize over any other parameters (in other words, if all are assumed to be known) then the minimal expected error is $\theta=1/\sqrt{F_{ii}}$. The inverse of the Fisher matrix provides an estimate of the parameter covariance matrix. Its diagonal elements  are the squares of the uncertainties in each parameter marginalizing over the others, while the off-diagonal terms yield the correlation coefficients between parameters. Note that the marginalized uncertainty is always greater than (or at most equal to) the non-marginalized one: marginalization can't decrease the error, and only has no effect if all other parameter are uncorrelated with it. It is also useful to define a Figure of Merit (denoted FoM for brevity in the results section) for each pair of parameters \cite{FMA2} which is the inverse of the area of their one-sigma confidence ellipse: a small area (meaning small uncertainties in the parameters) corresponds to a large figure of merit.

Previously known uncertainties on the parameters, known as priors, can be trivially added to the calculated Fisher matrix. This is manifestly the case for us: a plethora of standard cosmological datasets provide priors on our previously defined cosmological parameters $(\Omega_m,w_0,w_a)$, while local constraints on the  E\"{o}tv\"{o}s parameter $\eta$ from torsion balance and lunar laser ranging experiments \cite{Torsion,Lunar} provide priors on the dimensionless coupling $\zeta$. Specifically, we will assume the following fiducial values and prior uncertainties for our cosmological parameters
\be
\Omega_{m,fid}=0.3\,, \quad \sigma_{\Omega_m}=0.03
\ee
\be
w_{0,fid}=-0.9\,, \quad \sigma_{w_0}=0.1
\ee
\be
w_{a,fid}=0.3\,, \quad \sigma_{w_a}=0.3\,,
\ee
while for the coupling $\zeta$ we will consider three different scenarios
\be
\zeta_{fid}=0\,,\quad \zeta_{fid}=5\times10^{-7}\,,\quad \zeta_{fid}=5\times10^{-6}\,,
\ee
always with the same prior uncertainty
\be
\sigma_\zeta=10^{-4}\,.
\ee

Thus we will consider both the case where there are no $\alpha$ variations ($\zeta=0$), and the case where they exist: the case $\zeta=5\times10^{-6}$ corresponds to a coupling which saturates constraints from current data \cite{Pinho2,Pinho3}, while $\zeta=5\times10^{-7}$ illustrates an intermediate scenario.

The first ESPRESSO measurements of $\alpha$ should be obtained in the context of the consortium's Guaranteed Time Observations (GTO). The target list for these measurements has recently been selected: full details can be found in \cite{Masters}. Bearing this in mind we have studied the following three scenarios:
\begin{itemize}
\item {\it ESPRESSO Baseline}: we assumed that each of the targets on the list can be measured by ESPRESSO with an uncertainty of $\sigma_{\Delta\alpha / \alpha}=0.6\times10^{-6}$; this represents what we can currently expect to achieve on a time scale of 3-5 years (though this expectation needs to be confirmed at the time of commissioning of the instrument);
\item {\it ESPRESSO Ideal}: in this case we assumed a factor of three improvement in the uncertainty, $\sigma_{\Delta\alpha / \alpha}=0.2\times10^{-6}$; this represents somewhat optimistic uncertainties. This provides a useful comparison point, but in any case such an improved uncertainty should be achievable with additional integration time;
\item {\it ELT-HIRES}: We will also provide forecasts for a longer-term dataset, on the assumption that the same targets can be observed with the ELT-HIRES spectrograph \cite{HIRES}; in this case we assume an improvement in sensitivity by a factor of six relative to the ESPRESSO baseline scenario, coming from the larger collecting area of the telescope and additional improvements at the level of the spectrograph. Although at present not all details of the instrument and the telescope have been fixed, this is representative of the expected sensitivity of measurements on a 10-15 year time scale.
\end{itemize}

We note that our choices of possible theoretical and observational parameters span a broad range of possible scenarios. As a simple illustration of this point, let us consider a single measurement of $\alpha$ at redshift $z=2$. In the case of the dilaton model we have the simple relation $\Delta\alpha/\alpha(z=2)\sim0.5\zeta$. Thus if $\zeta=5\times10^{-7}$ a single precise and accurate measurement of $\alpha$ with ESPRESSO baseline sensitivity would not detect its variation, while ELT-HIRES would detect it at 2.5 standard deviations. On the other hand, for  $\zeta=5\times10^{-6}$ (which as previously mentioned saturates current bounds) a single $z=2$ ESPRESSO baseline measurement would detect a variation at $4\sigma$ and ELT-HIRES would detect it at $25\sigma$.

Before proceeding with our general analysis it is instructive to provide a simple analytic illustration for the dilaton model, in which case the $\alpha$ variation is given by Eq. \ref{dilalpha}. Let's further assume that $\Omega_\phi$ (or equivalently $\Omega_m$) is perfectly known, so we are left with a two-dimensional parameter space $(\zeta,w_0)$. Including priors on both $\zeta$ and $w_0$ (respectively denoted $\sigma_\zeta$ and $\sigma_w$), the Fisher matrix is
\be
[F(\zeta,w_0)]=\begin{bmatrix}
Q^2(1+w_0)+\frac{1}{\sigma^2_\zeta} \hfill & \frac{1}{2}Q^2\zeta \hfill \\
\frac{1}{2}Q^2\zeta \hfill & \frac{Q^2\zeta^2}{4(1+w_0)}+\frac{1}{\sigma^2_w} \hfill
\end{bmatrix}\,,
\ee
where we have defined
\be
Q^2=3\Omega_\phi\sum_i\left[\frac{\log(1+z_i)}{\sigma_{\alpha i}}\right]^2\,.
\ee
The un-marginalized uncertainties are
\be
\theta_\zeta=\frac{\sigma_\zeta}{\sqrt{1+(1+w_0)Q^2\sigma^2_\zeta}}
\ee
\be
\theta_w=\frac{\sigma_w}{\sqrt{1+\frac{\zeta^2}{4(1+w_0)}Q^2\sigma^2_w}}
\ee
while the determinant of F is
\be
det F=Q^2\left[\frac{1+w_0}{\sigma^2_w}+\frac{\zeta^2}{4(1+w_0)\sigma^2_\zeta} \right]+\frac{1}{\sigma^2_w\sigma^2_\zeta}\,;
\ee
this would be zero in the absence of priors---a point already discussed in \cite{Erminia2}---but as mentioned above cosmological data and local tests of the WEP do provide us with these priors. As expected, if $\zeta=0$ the two parameters decorrelate, and there is no new information on the equation of state ($\theta_w=\sigma_w$): if $\zeta=0$ we will always measure $\Delta\alpha/\alpha=0$ regardless of the experimental sensitivity.

Now we can calculate the covariance matrix
\be
[C(\zeta,w_0)]=\frac{1}{det F}\begin{bmatrix}
\frac{Q^2\zeta^2}{4(1+w_0)}+\frac{1}{\sigma^2_w} \hfill & -\frac{1}{2}Q^2\zeta \hfill \\
-\frac{1}{2}Q^2\zeta \hfill & Q^2(1+w_0)+\frac{1}{\sigma^2_\zeta} \hfill
\end{bmatrix}\,,
\ee
and the correlation coefficient $\rho$ can be written
\be
\rho=\left[1+\frac{4(1+w_0)}{Q^2\zeta^2\sigma^2_w} +\frac{1}{(1+w_0)Q^2\sigma^2_\zeta}+\frac{4}{\zeta^2Q^4\sigma^2_\zeta\sigma^2_w} \right]^{-1/2}\,.
\ee
We thus confirm the physical intuition that in the limit $\zeta\longrightarrow0$, the two parameters become independent ($\rho\to0$). The general marginalized uncertainties are
\be
\frac{1}{\sigma^2_{\zeta,new}}=\frac{1}{\sigma^2_\zeta}+\frac{1}{\sigma^2_w}\frac{(1+w_0)Q^2}{\frac{\zeta^2Q^2}{4(1+w_0)}+\frac{1}{\sigma^2_w}}
\ee
\be
\frac{1}{\sigma^2_{w,new}}=\frac{1}{\sigma^2_w}+\frac{1}{\sigma^2_\zeta}\frac{\zeta^2Q^2}{4(1+w_0)}\frac{1}{(1+w_0)Q^2+\frac{1}{\sigma^2_\zeta}}\,;
\ee
In the particular case where the fiducial model is $\zeta=0$ the former becomes
\be
\frac{1}{\sigma^2_{\zeta,new}}=\frac{1}{\sigma^2_\zeta}+(1+w_0)Q^2
\ee
while the latter trivially gives $\sigma_{w,new}=\sigma_w$. As previously mentioned, we have used these analytic results to validate our more generic numerical code (where furthermore $\Omega_m$ will also be allowed to vary).

\begin{table*}[ht]
\begin{center}
\caption{The first three lines show the one sigma forecasted uncertainties on the dimensionless coupling parameter $\zeta$, marginalizing over the remaining model parameters, for the various choices of fiducial cosmological model and dataset of $\alpha$ measurements. The fiducial value of the coupling is $\zeta_{fid}=0$ in all cases. The last line shows the corresponding one-sigma uncertainty on the E\"{o}tv\"{o}s parameter $\eta$, in the least constraining case of the $w_0=const.$ model }
\label{tablezero}
\begin{tabular}{| c | c | c | c |}
\hline
Model & ESPRESSO baseline & ESPRESSO ideal & ELT-HIRES \\
\hline
$w_0=const.$ & $4.6\times10^{-7}$ & $1.5\times10^{-7}$ & $7.6\times10^{-8}$ \\
Dilaton & $3.2\times10^{-7}$ & $1.1\times10^{-7}$ & $5.3\times10^{-8}$  \\
CPL & $3.1\times10^{-7}$ & $1.0\times10^{-7}$ & $5.1\times10^{-8}$  \\
\hline
$\eta$ & $2.1\times10^{-16}$ & $2.3\times10^{-17}$ & $5.8\times10^{-18}$ \\
\hline
\end{tabular}
\end{center}
\end{table*}
\begin{figure}
\begin{center}
\includegraphics[width=3.3in,keepaspectratio]{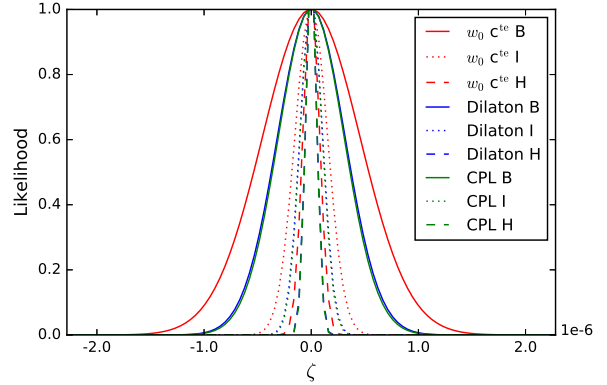}
\caption{\label{fig1}One sigma forecasted uncertainties on $\zeta$, marginalizing over the remaining model parameters, for the various choices of fiducial cosmological model (shown in different colors) and dataset of $\alpha$ measurements (with solid, dotted and dashed lines respectively depicting ESPRESSO baseline, ESPRESSO ideal and ELT-HIRES, cf. the main text). The fiducial value of the coupling is $\zeta_{fid}=0$ in all cases.}
\end{center}
\end{figure}

\section{Results}

We start with the case where there is no coupling between the scalar field and the electromagnetic sector of the theory, such that  $\zeta_{fid}=0$ . We emphasize that if the cosmological model Lagrangian does contain a dynamical scalar field, the suppression of such a coupling will require a (still unknown) symmetry \cite{Carroll,Dvali,Chiba}. In this case precise $\alpha$ measurements will find null results which can be translated into bounds on $\zeta$, whose one-sigma uncertainties, marginalized over $\Omega_m$, $w_0$ and (for the case of the CPL model), $w_a$, are displayed in Table \ref{tablezero} and in Fig. \ref{fig1}.

For comparison, the current two-sigma bound on $\zeta$ is $|\zeta|<5\times10^{-6}$, with a mild dependence on the choice of fiducial dark energy model \cite{Pinho2,Pinho3}. Thus in this case we expect ESPRESSO to improve current bounds on $\zeta$ by about one order of magnitude. Naturally these improvements also lead to stronger bounds on the  E\"{o}tv\"{o}s parameter: we note that constraints from ESPRESSO should be stronger than those expected from the ongoing tests with the MICROSCOPE satellite \cite{MICROSCOPE}, while those from ELT-HIRES should be competitive with those of the proposed STEP satellite \cite{STEP} (though at present the wavelength coverage and sensitivity of the latter are relatively uncertain).

Table \ref{tablezero} also shows that there is a mild dependence on the choice of underlying dark energy model. This has been previously studied, and is well understood---refer to \cite{Pinho2,Pinho3,Erminia2} for further discussion of this point. The dilaton model is a 'freezing' dark energy model. Thus, according to Eq. \ref{eq:dalfa}, a dilaton model with a given value of $w_0$ will have a value of $\Delta\alpha/\alpha(z)$ that is larger than the corresponding value for a model with a constant equation of state with the same value of $w_0$. Thus, for similar cosmological priors, null measurements of $\alpha$ will provide slightly stronger constraints for the dilaton case. The same argument applies for the CPL case, where the additional free parameter $w_a$ further enlarges the range of possible values of $\alpha$.

\begin{figure*}
\begin{center}
\includegraphics[width=3.2in,keepaspectratio]{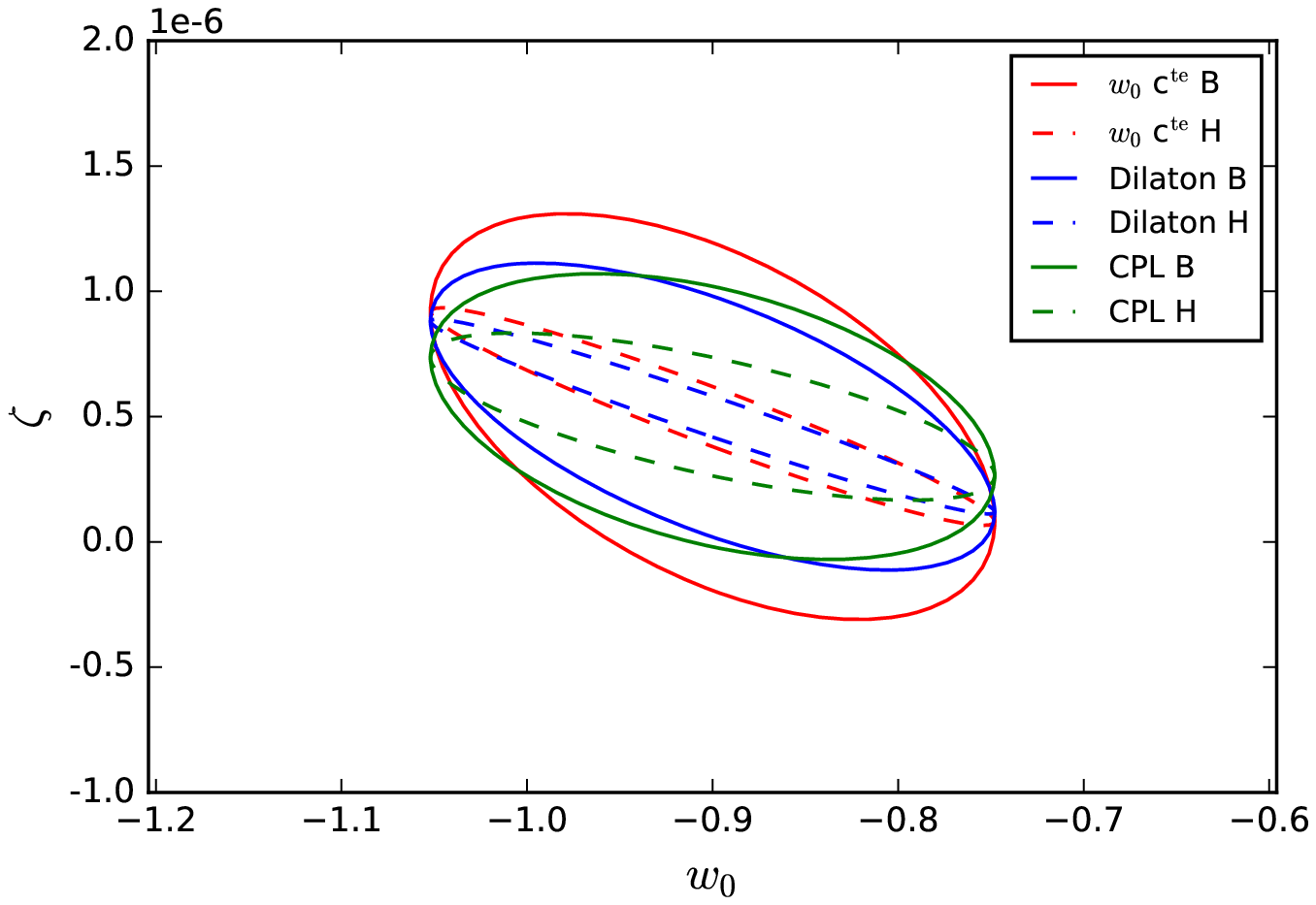}
\includegraphics[width=3.2in,keepaspectratio]{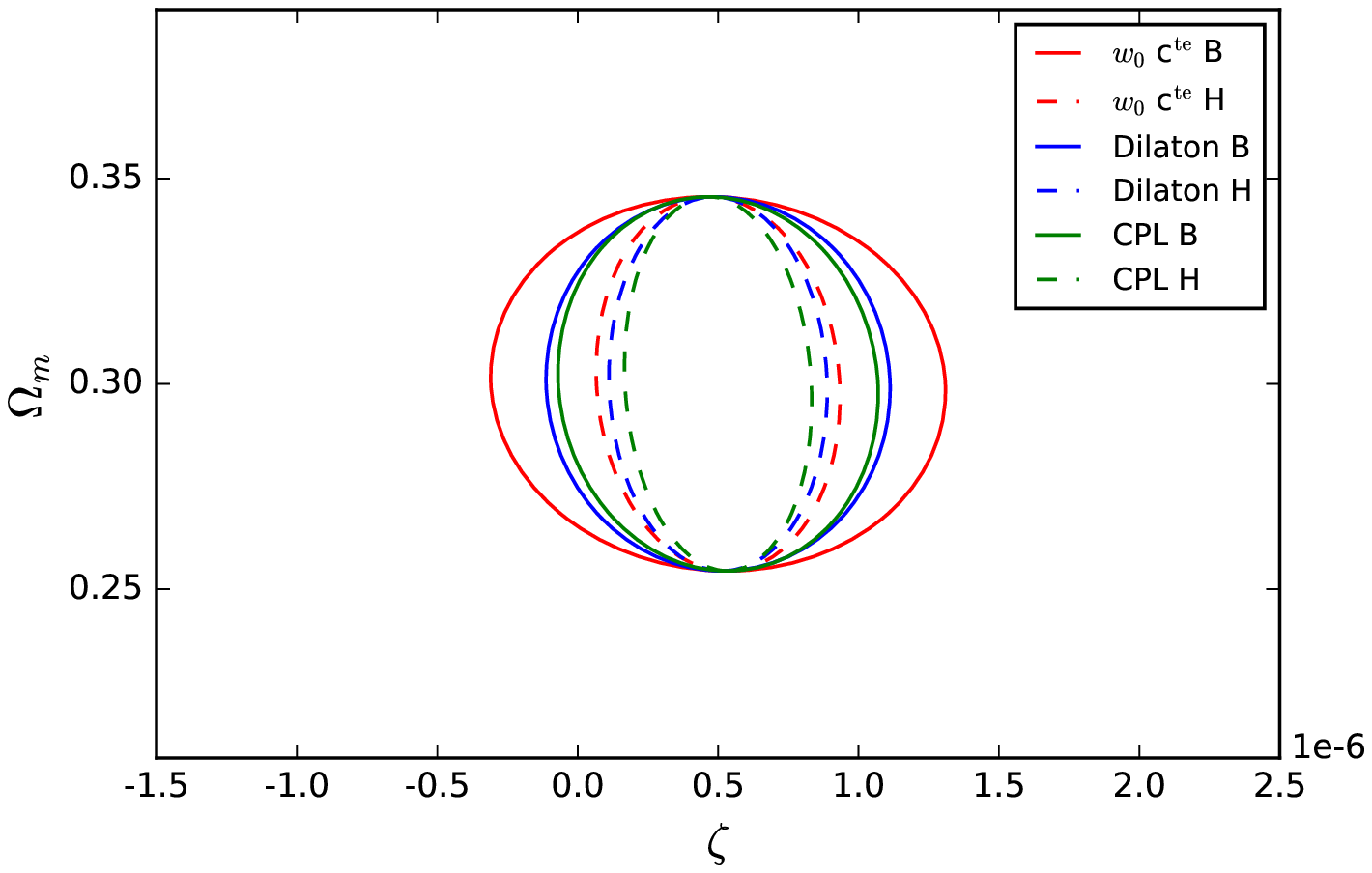}
\end{center}
\caption{\label{fig2}Forecasted uncertainties in the $\zeta-w_0$ and $\zeta-\Omega_m$ planes (left and right panels), marginalizing over the remaining model parameters, for the various choices of fiducial cosmological model (shown in different colors) and dataset of $\alpha$ measurements (solid lines for the ESPRESSO baseline case and dashed lines for ELT-HIRES, cf. the main text), for a fiducial value of the coupling $\zeta=5\times10^{-7}$.}
\end{figure*}

\begin{figure*}
\begin{center}
\includegraphics[width=3.2in,keepaspectratio]{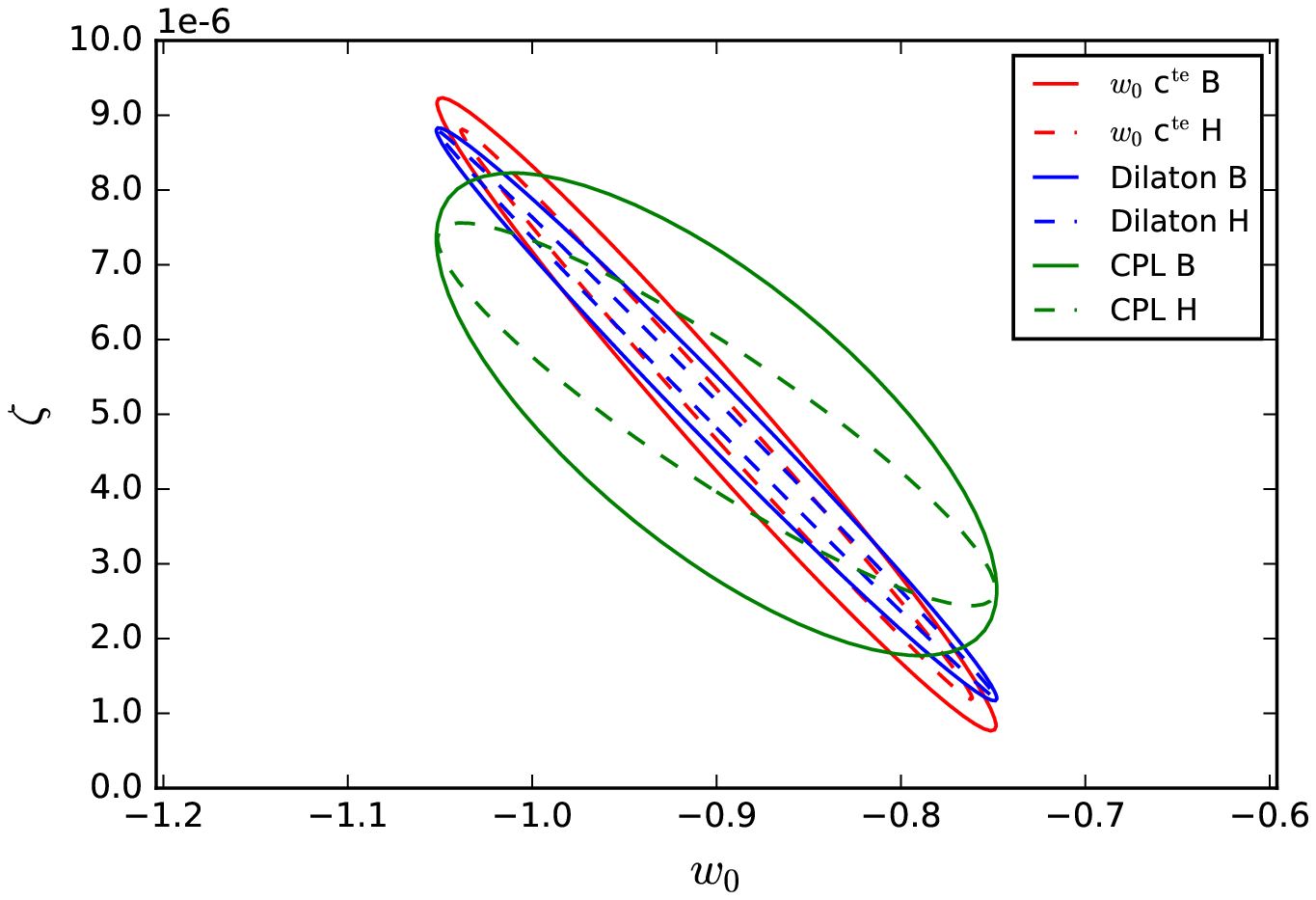}
\includegraphics[width=3.2in,keepaspectratio]{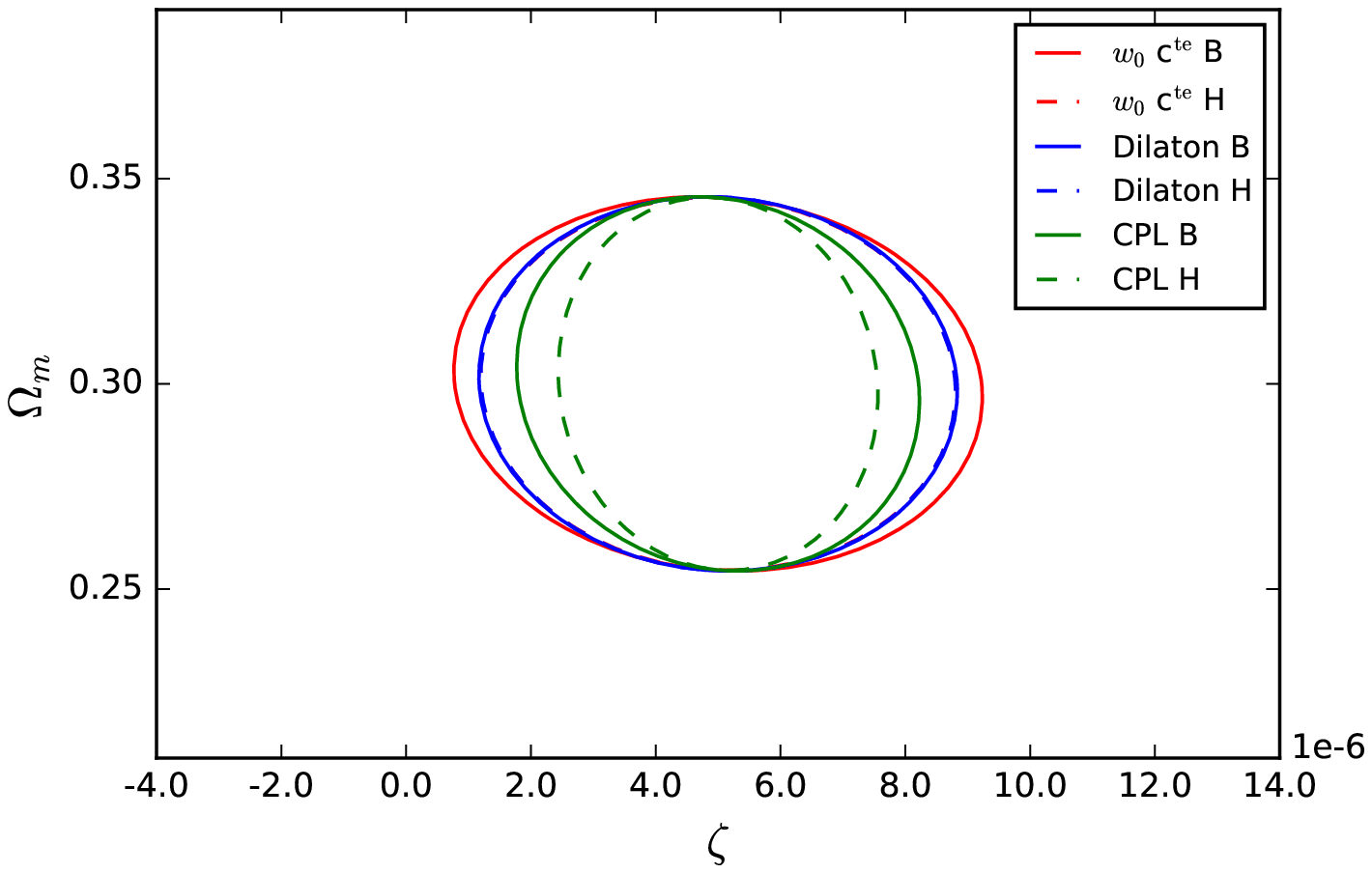}
\end{center}
\caption{\label{fig3}Same as Fig. \protect\ref{fig2}, for a fiducial values of the coupling $\zeta=5\times10^{-6}$.}
\end{figure*}

\begin{figure*}
\begin{center}
\includegraphics[width=3.2in,keepaspectratio]{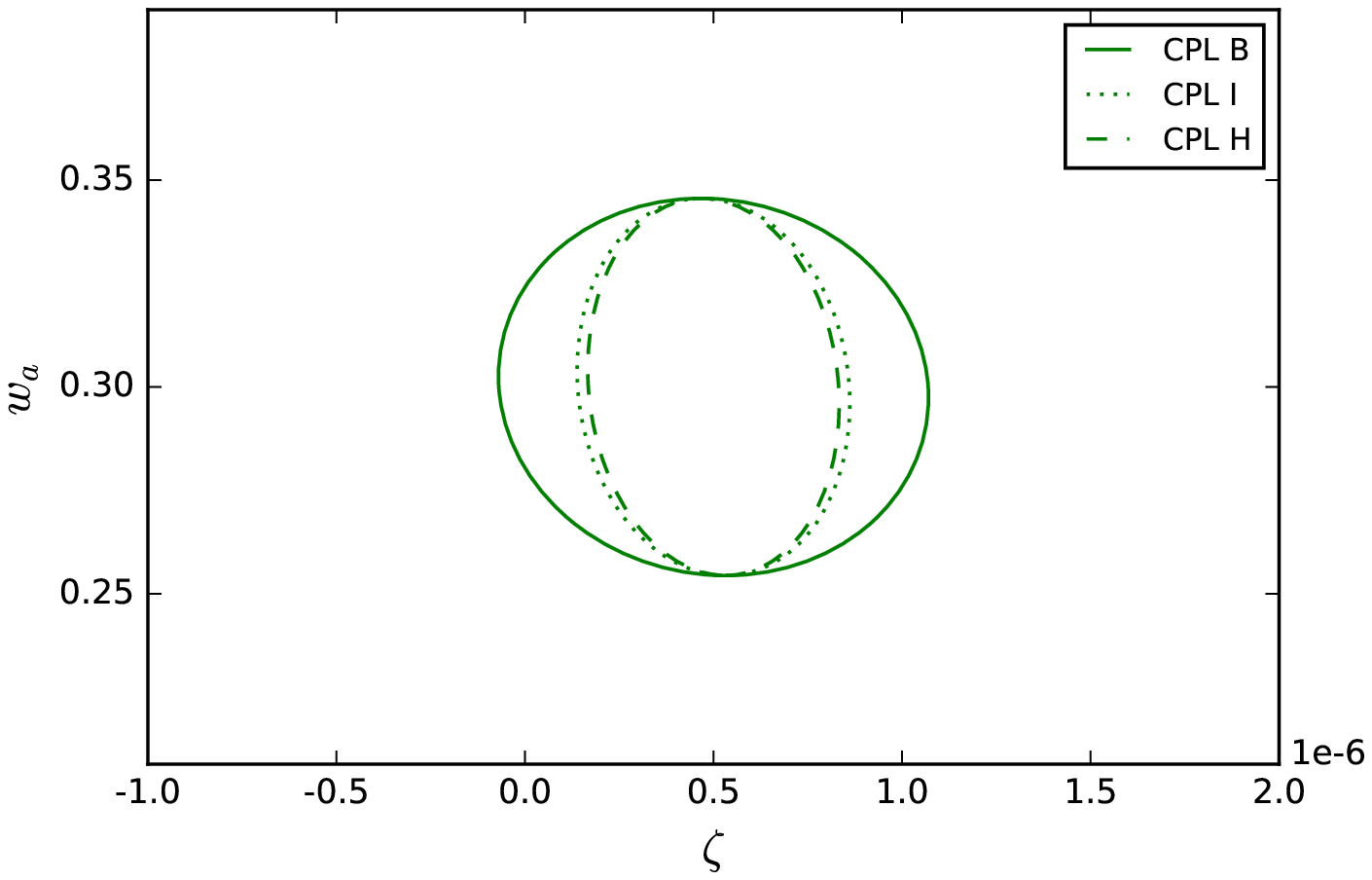}
\includegraphics[width=3.2in,keepaspectratio]{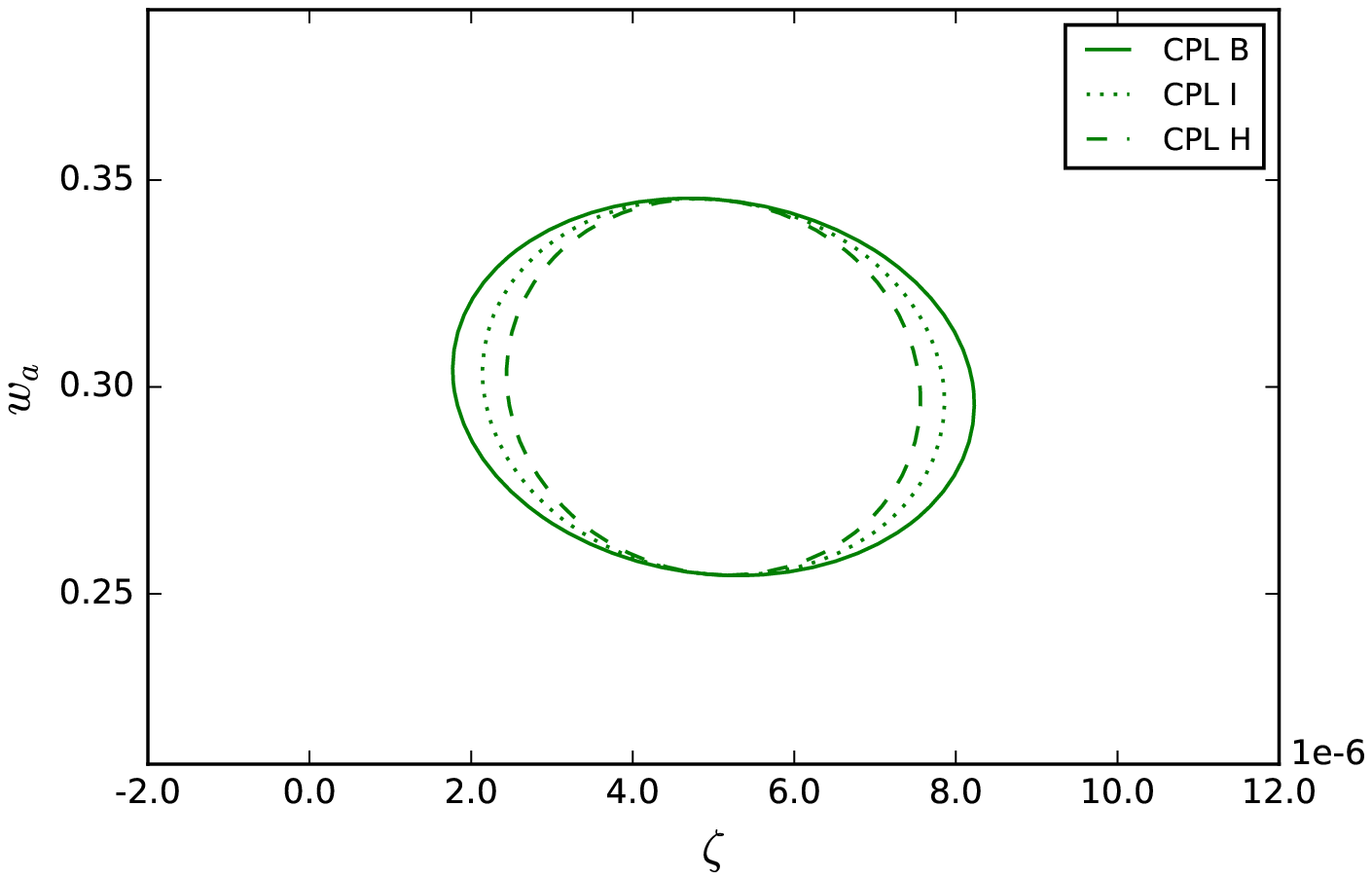}
\end{center}
\caption{\label{fig4}Forecasted uncertainties in the $\zeta-w_a$ plane, marginalizing over the remaining model parameters, for the CPL model, with various choices of fiducial cosmological dataset of $\alpha$ measurements (with solid, dotted and dashed lines respectively depicting ESPRESSO baseline, ESPRESSO ideal and ELT-HIRES, cf. the main text), for fiducial values of the coupling $\zeta=5\times10^{-7}$ (left panel) or $\zeta=5\times10^{-6}$ (right panel) .}
\end{figure*}

Now we consider the case where an $\alpha$ variation does exist, corresponding to a non-zero fiducial value of the dimensionless coupling $\zeta$. In this case the marginalized sensitivity on the parameter $\zeta$ will be weakened due to its correlations with other parameters. On the other hand, the $\alpha$ measurements can themselves help in constraining the cosmological parameters. The results of our analysis are summarized in Tables \ref{tableconst}, \ref{tabledil} and \ref{tablecpl}, respectively for the constant equation of state, dilaton and CPL fiducial models, and also shown in Figs. \ref{fig2}, \ref{fig3} and \ref{fig4}.

\begin{table*}
\caption{Results of the Fisher Matrix analysis for the case of the constant dark energy equation of state model. The first three lines show the correlation coefficients $\rho$ for each pair of parameters, the following three the Figure of Merit for each pair of parameters, and the last two the one-sigma marginalized uncertainties for the coupling $\zeta$ and the present-day value of the dark energy equation of state $w_0$.}
\label{tableconst}
\begin{center}
\begin{tabular}{| c | c c | c c | c c |} 
\hline
{ } & ESPRESSO & baseline &  ESPRESSO & ideal & ELT-HIRES  & { } \\
Parameter & $\zeta=5\times10^{-7}$ & $\zeta=5\times10^{-6}$ & $\zeta=5\times10^{-7}$ & $\zeta=5\times10^{-6}$ & $\zeta=5\times10^{-7}$ & $\zeta=5\times10^{-6}$   \\
\hline
$\rho(\zeta,w_0)$      & -0.516 & -0.984 & -0.873 & -0.995 & -0.961 & -0.996 \\
$\rho(\Omega_m,w_0)  $ & $-1.3\times10^{-5}$ & $-1.3\times10^{-3}$ & $-1.1\times10^{-4}$ & $-1.1\times10^{-2}$ & $-4.5\times10^{-4}$ & $-4.1\times10^{-2}$ \\
$\rho(\zeta,\Omega_m)$ & -0.039 & -0.074 & -0.066 & -0.067 & -0.073 & -0.042 \\
\hline
$FoM(\zeta,w_0)/10^6$      & 3.03 & 2.77 & 9.02 & 5.49 & 17.6 & 6.85 \\
$FoM(\Omega_m,w_0)$   & 46.1 & 46.3 & 46.1 & 47.4 & 46.2 & 50.9 \\
$FoM(\zeta,\Omega_m)/10^6$ & 8.66 & 1.66 & 14.7 & 1.72 & 16.2 & 1.85 \\
\hline
$\sigma(\zeta)$ & $5.3\times10^{-7}$ & $2.8\times10^{-6}$ & $3.1\times10^{-7}$ & $2.7\times10^{-6}$ & $2.9\times10^{-7}$ & $2.5\times10^{-6}$ \\
$\sigma(w_0)$   & 0.100 & 0.100 & 0.100 & 0.098 & 0.100 & 0.091 \\
\hline
\end{tabular}
\end{center}
\end{table*}
\begin{table*}
\caption{Same as Table \protect\ref{tableconst} for the case of the dilaton model.}
\label{tabledil}
\begin{center}
\begin{tabular}{| c | c c | c c | c c |}
\hline
{ } & ESPRESSO & baseline &  ESPRESSO & ideal & ELT-HIRES  & { } \\
Parameter & $\zeta=5\times10^{-7}$ & $\zeta=5\times10^{-6}$ & $\zeta=5\times10^{-7}$ & $\zeta=5\times10^{-6}$ & $\zeta=5\times10^{-7}$ & $\zeta=5\times10^{-6}$   \\
\hline
$\rho(\zeta,w_0)$      & -0.620 & -0.991 & -0.921 & -0.998 & -0.978 & -0.999 \\
$\rho(\Omega_m,w_0)  $ & $-2.7\times10^{-7}$ & $-2.7\times10^{-5}$ & $-2.7\times10^{-7}$ & $-2.7\times10^{-5}$ & $-2.7\times10^{-7}$ & $-2.7\times10^{-5}$ \\
$\rho(\zeta,\Omega_m)$ & -0.027 & -0.042 & -0.039 & -0.043 & -0.042 & 0.043 \\
\hline
$FoM(\zeta,w_0)/10^6$      & 4.38 & 4.15 & 13.1 & 9.21 & 25.7 & 11.6 \\
$FoM(\Omega_m,w_0)$   & 46.1 & 46.1 & 46.1 & 46.1 & 46.1 & 46.1 \\
$FoM(\zeta,\Omega_m)/10^6$ & 11.4 & 1.83 & 17.0 & 1.84 & 18.1 & 1.85 \\
\hline
$\sigma(\zeta)$ & $4.0\times10^{-7}$ & $2.5\times10^{-6}$ & $2.7\times10^{-7}$ & $2.5\times10^{-6}$ & $2.6\times10^{-7}$ & $2.5\times10^{-6}$ \\
$\sigma(w_0)$   & 0.100 & 0.100 & 0.100 & 0.100 & 0.100 & 0.100 \\
\hline
\end{tabular}
\end{center}
\end{table*}
\begin{table*}
\caption{Results of the Fisher Matrix analysis for the case of the CPL parametrization. The first six lines show the correlation coefficients $\rho$ for each pair of parameters, the following six the Figure of Merit for each pair of parameters, and the last three the one-sigma marginalized uncertainties for the coupling $\zeta$ and the dark energy equation of state parameters $w_0$ and $w_a$.}
\label{tablecpl}
\begin{center}
\begin{tabular}{| c | c c | c c | c c |}
\hline
{ } & ESPRESSO & baseline &  ESPRESSO & ideal & ELT-HIRES  & { } \\
Parameter & $\zeta=5\times10^{-7}$ & $\zeta=5\times10^{-6}$ & $\zeta=5\times10^{-7}$ & $\zeta=5\times10^{-6}$ & $\zeta=5\times10^{-7}$ & $\zeta=5\times10^{-6}$   \\
\hline
$\rho(\zeta,w_0)$      & -0.412 & -0.728 & -0.650 & -0.822 & -0.705 & -0.914 \\
$\rho(\Omega_m,w_0)$   & $1.6\times10^{-7}$ & $1.6\times10^{-5}$ & $4.0\times10^{-6}$ & $3.3\times10^{-4}$ & $1.7\times10^{-5}$ & $1.2\times10^{-3}$ \\
$\rho(w_0,w_a)$        & $6.2\times10^{-9}$ & $4.6\times10^{-7}$ & $1.8\times10^{-5}$ &$ 1.3\times10^{-3}$ & $7.9\times10^{-5}$ & $3.4\times10^{-3}$ \\
$\rho(\zeta,\Omega_m)$ & -0.057 & -0.095 & -0.089 & -0.080 & -0.095 & -0.067 \\
$\rho(\zeta,w_a)$      & -0.395 & -0.663 & -0.620 & -0.557 & -0.663 & -0.387 \\
$\rho(\Omega_m,w_a)$   & $-9.3\times10^{-5}$ & $-8.9\times10^{-3}$ & $-8.4\times10^{-4}$ & $-6.0\times10^{-2}$ & $-3.3\times10^{-3}$ & $-1.5\times10^{-1}$ \\
\hline
$FoM(\zeta,w_0)/10^6$      & 4.05 & 0.950 & 7.65 & 1.29 & 8.89 & 2.04 \\
$FoM(\Omega_m,w_0)$   & 46.1 & 46.2 & 46.1 & 46.3 & 46.1 & 46.5 \\
$FoM(w_0,w_a)$        & 4.62 & 4.86 & 4.64 & 6.47 & 4.70 & 10.1 \\
$FoM(\zeta,\Omega_m)/10^6$ & 12.3 & 2.18 & 19.5 & 2.47 & 21.1 & 2.75 \\
$FoM(\zeta,w_a)/10^6$      & 1.34 & 0.305 & 2.48 & 0.414 & 2.86 & 0.649 \\
$FoM(\Omega_m,w_a)$   & 15.4 & 16.2 & 15.5 & 21.6 & 15.7 & 34.1 \\
\hline
$\sigma(\zeta)$ & $3.8\times10^{-7}$ & $2.1\times10^{-6}$ & $2.4\times10^{-7}$ & $1.9\times10^{-6}$ & $2.2\times10^{-7}$ & $1.7\times10^{-6}$ \\
$\sigma(w_0)$   & 0.100 & 0.100 & 0.100 & 0.100 & 0.100 & 0.100 \\
$\sigma(w_a)$   & 0.300 & 0.285 & 0.299 & 0.214 & 0.294 & 0.137 \\
\hline
\end{tabular}
\end{center}
\end{table*}

Starting with the constant equation of state and dilaton models, we confirm the strong anticorrelation between $\zeta$ and $w_0$ (which naturally is weaker for smaller values of the coupling): since the $\alpha$ variation depends both on the strength of the coupling and on how fast the scalar field is moving---which depends on (1+w(z)), cf. Eq. \ref{dynphi}---to a first approximation one can increase one and decrease the other and still get similar $\alpha$ variations. On the other hand, the present-day value of the matter density is not significantly correlated with the other parameters, as is clear from the right-hand side panels of Figs. \ref{fig2} and \ref{fig3}.

Comparing the two models the correlations are somewhat weaker in the dilaton case, thus leading to a better sensitivity on the coupling $\zeta$. Overall, with the range of assumed couplings the ESPRESSO GTO measurements would detect a non-zero $\zeta$ at between one and two standard deviations, while the same observations with the foreseen ELT-HIRES would ensure a two-sigma detection. We also note that for the largest permissible values of the coupling, ELT-HIRES measurements can improve constraints on the dark energy equation of state $w_0$ by up to ten percent.

The case of the CPL parametrization is particularly illuminating. Here the behavior of the dark energy equation of state depends on two parameters, $w_0$ and $w_a$ (while in the case of the other models we assumed that it just depended on the former). Each of these parameters is still anticorrelated with $\zeta$, for the reasons already explained, but this anticorrelation is now weaker than in the dilaton or the constant equation of state cases, enabling stronger constraints on $\zeta$. This is manifest in the left-hand side panels of Figs. \ref{fig2} and \ref{fig3}. Thus in the case of the largest currently allowed value $\zeta=5\times10^{-6}$ ELT-HIRES observations of the ESPRESSO GTO sample would detect a non.zero $\zeta$ at the $99.7\%$ (3$\sigma$) confidence level.

It is particularly worthy of note that the two dark energy equation of state parameters are not significantly correlated. This occurs because measurements of $\alpha$ typically span a sufficiently large redshift range (in our case roughly $1<z<3$) to make the roles of both in the redshift dependence of $\alpha$ sufficiently distinct. The practical result of this is that in the case of large values of $\zeta$ these measurements can significantly improve constraints on $w_a$---by more than a factor of two for the case of ELT-HIRES, and by about $30\%$ for ESPRESSO ideal data, in the case of a large coupling---see the last line in Table \ref{tablecpl}. Thus $\alpha$ measurements can ideally complement cosmological probes in mapping the behavior of dynamical dark energy.


\section{Discussion and conclusions}

We have used standard Fisher Matrix analysis techniques to study the cosmological impact of short and medium-term astrophysical tests of the stability of the fine-structure constant. The ESPRESSO spectrograph will be commissioned at the VLT in late 2017, and since it will be located at the combined Coud\'e focus it will be able to incoherently combine light from the four VLT unit telescopes. On the other hand, the European Extremely Large Telescope, with first light expected in 2024, will have a 39.3m primary mirror. The larger telescope collecting areas are one of the reasons behind the expected improvements in the sensitivity of these measurements (which are photon-starved). The other such reason pertains to technological improvements in the spectrographs themselves, enabling, among others, higher resolution and stability \cite{ESPRESSO,HIRES}.

Our analysis demonstrates that whether these measurements lead to detections of variations or to improved null results, they will have important implications for cosmology as well as for fundamental physics. In the scenario where there are no $\alpha$ variations, ESPRESSO can improve current bounds on WEP violations by up to two orders of magnitude: such bounds would be stronger bounds than those expected from the MICROSCOPE satellite. Similarly, constraints from the high-resolution spectrograph at the E-ELT should be competitive with those of the proposed STEP satellite (although in this case one should be mindful of the caveat that both facilities are currently still in early stages of development).

In the opposite case where an $\alpha$ variation is detected, and quite apart from the direct implications (direct evidence of Einstein Equivalence Principle violation, falsifying the notion of gravity as a purely geometric phenomenon, and of a fifth interaction in nature \cite{Uzan}) there are additional implications for cosmology. While the anticorrelations between the scalar field electromagnetic coupling $\zeta$ and the dark energy equation of state parameters mean that constraints on $\zeta$ will in this case be weaker than in the null case, the new facilities will extend the range of couplings that can be meaningfully probed by at least one order of magnitude. Moreover, these measurements are particularly sensitive to the dynamics of dark energy, and could conceivably improve constraints on $w_a$ by more than a factor of two.

We emphasize that the analysis we have presented is conservative in at least one sense: our sample of $\alpha$ measurements consisted only of the 14 measurements in the range $1<z<3$ foreseen for the fundamental physics part of the ESPRESSO GTO \cite{Leite2,Masters}. This is to be compared to the 293 archival measurements of Webb {\it et al.} \cite{Dipole}, in the approximate redshift range $0.5<z<4.2$. The latter contains data gathered over a period of about ten years from two of the world's largest telescopes, while the 14 GTO targets were chosen on the grounds that they are the best currently known targets for these measurements (and are visible from the location of the VLT, at Cerro Paranal in Chile) and improving the measurements on these targets will have a significant impact in the field. Nevertheless it is clear that in a time scale of 5-10 years a significantly larger dataset could be obtained, leading to even stronger constraints.

Although our analysis focused on time (redshift) variations of $\alpha$, the 14 ESPRESSO GTO measurements will also test possible spatial variations. In particular, if one assumes that $\alpha$ varies as a pure spatial dipole with the best-fit parameters given in the analysis of \cite{Dipole}, then along the 14 ESPRESSO GTO lines of sight one expects variations ranging between ${\Delta\alpha / \alpha}=-3.8\times10^{-6}$ and ${\Delta\alpha / \alpha}=+4.1\times10^{-6}$. Moreover, in 11 of the 14 lines of sight those best-fit parameters predict variations whose absolute value is larger than the nominal statistical uncertainty of the ESPRESSO baseline measurements, $\sigma_{\Delta\alpha / \alpha}=0.6\times10^{-6}$. Thus the measurements will provide a test of such spatial variations at a high level of statistical significance, at least at the claimed level of parts-per-million in amplitude.

Finally, although our constraints are mildly dependent on the choice of fiducial cosmological model---a result that confirms analyses of current data \cite{Pinho2,Pinho1,Pinho3}---this is actually a desirable feature. Should ESPRESSO or ELT-HIRES detect $\alpha$ variations, these measurements will ideally complement other canonical observables in selecting between otherwise indistinguishable cosmological models. A more detailed study of this procedure is left for subsequent work.

\section*{Acknowledgements}
We are grateful to Ana Marta Pinho and Paolo Molaro for helpful discussions on the subject of this work. This work was done in the context of project PTDC/FIS/111725/2009 (FCT, Portugal), with additional support from grant UID/FIS/04434/2013. ACL is supported by an FCT fellowship (SFRH/BD/113746/2015), under the FCT PD Program PhD::SPACE (PD/00040/2012), and by the Gulbenkian Foundation through \textit{Programa de Est\'{i}mulo \`{a} Investiga\c{c}\~{a}o 2014}, grant number 2148613525. CJM is supported by an FCT Research Professorship, contract reference IF/00064/2012, funded by FCT/MCTES (Portugal) and POPH/FSE (EC).

\bibliographystyle{model1-num-names}
\bibliography{alpha}
\end{document}